\documentclass[aps,prl,twocolumn,notitlepage]{revtex4-2}

\usepackage{graphicx} 
\usepackage{lipsum}
\usepackage{amsmath,amssymb,graphicx,color}
\usepackage{bm}
\usepackage{natbib}
\usepackage{placeins}
\setcitestyle{square}
\usepackage{braket}
\usepackage{makerobust}
\usepackage{xcolor}
\usepackage{hyperref}

\hypersetup{
 colorlinks,
 breaklinks=true,
 plainpages=false,
 citecolor=blue,
 linkcolor=blue,
 urlcolor=blue,
 bookmarksopen=true,
 bookmarksnumbered=false,
 bookmarksdepth=5
}

\newcommand{\makeauthor}[2]{\newcommand{#1}[1]{{%
 \sffamily\color{#2}{%
 \bfseries\begingroup\escapechar=-1\edef\x{\endgroup\string#1}\x:%
 } ##1}}%
 \MakeRobustCommand#1}
\makeauthor{\dkm}{red}
\makeauthor{\rew}{blue}
\makeauthor{\mb}{cyan}
\makeauthor{\jb}{purple}

\begin{document}

\title{Braiding of Majorana Zero Modes in Vortex Cores}
\author{Maxwell Buss, Jasmin Bedow, Prashant Gupta, and Dirk K. Morr}
\affiliation{University of Illinois Chicago, Chicago, IL 60607, USA}
\date{\today}

\begin{abstract}
We demonstrate the successful simulation of $\sqrt{Z}$-, $\sqrt{X}$- and $X$-quantum gates using Majorana zero modes (MZMs) that emerge in magnetic vortices located in topological superconductors.  We compute the transition probabilities and geometric phase differences accounting for the full many-body dynamics and show that qubit states can be read out by fusing the vortex core MZMs and measuring the resulting charge density. We visualize the gate processes using the time- and energy-dependent non-equilibrium local density of states. Our results demonstrate the feasibility of employing vortex core MZMs for the realization of fault-tolerant topological quantum computing.
\end{abstract}

\maketitle

Majorana zero modes (MZMs) offer a promising route towards the implementation of fault-tolerant quantum computing \cite{Nayak2008,Sarma2015}. These exotic particles, harboring the non-Abelian exchange statistics of Ising anyons, have been reported to exist at the ends of one-dimensional (1D) topological superconductors \cite{Nadj-Perge2014,Ruby2015,Pawlak2016,Kim2018} as well as defects in two-dimensional (2D) topological superconductors \cite{Menard2019,LoConte2024,Wang2018,Machida2019,Kong2019,Zhu2020}. Numerous approaches to realizing topological quantum gates using Majorana zero modes have been proposed \cite{Alicea2011,Halperin2012,Sekania2017,Harper2019,Tutschku2020,Tanaka2022,Mascot2023,Amorim2015,Kraus2013,Aasen2016,Karzig2017,Zhou2022,Hyart2013,Li2016,Sanno2021,Cheng2016,Chen2022,Wong2023,Bedow2024,Bedow2025}, which rely on the ability to dynamically manipulate the electronic structure at the atomic scale in order to spatially move MZMs, such as tuning the chemical potential \cite{Alicea2011,Halperin2012,Sekania2017,Harper2019,Tutschku2020,Tanaka2022,Mascot2023}, coupling constants \cite{Amorim2015,Kraus2013,Aasen2016,Karzig2017,Zhou2022}, magnetic fields \cite{Hyart2013,Li2016}, the phase of the superconducting order parameter \cite{Sanno2021} or the local magnetic structure \cite{Bedow2024,Bedow2025}. 
Alternative proposals \cite{Tewari2007, Wu2017, November2019, Beenakker2019, Posske2020, Ma2021} utilize the MZMs emerging in magnetic vortices that are located in 2D topological regions, such as magnet-superconductor hybrid (MSH) systems \cite{Menard2019,LoConte2024} or the iron-chalcogenide superconductor FeSe$_{1-x}$Te$_{x}$ \cite{Wang2018,Machida2019,Kong2019,Zhu2020}. While it has been suggested that these MZMs could be braided by dragging magnetic vortices using the tip of a scanning tunneling microscope (STM) \cite{Ge2016}, the cantilever of a magnetic force microscope \cite{Straver2008,November2019,Polshyn2019} or a superconducting quantum interference device (SQUID) \cite{Gardner2002,Kalisky2011,Kremen2016}, a dynamical simulation of these braiding processes and the resulting topologically protected quantum gates on experimentally relevant time and length scales has not been achieved yet. 

In this article, we demonstrate the successful simulation of $\sqrt{Z}$-, $\sqrt{X}$- and $X$-gates using vortex core MZMs. To this end, we compute the relevant transition probabilities and geometric phase differences accounting for the full many-body dynamics by using recent advances in simulating non-equilibrium phenomena in superconductors \cite{Mascot2023,Bedow2024,Hodge2025_2}. We show that qubit states can be read out by fusing the vortex core MZMs and measuring the resulting charge density. Moreover, we visualize the spectroscopic signatures of these processes using the time- and energy-dependent non-equilibrium local density of states (LDOS), which is proportional to the time-dependent differential conductance measured in STM experiments \cite{Bedow2022}. Our results demonstrate the feasibility of vortex core MZMs for the realization of fault-tolerant topological quantum computing in topological superconductors.

{\it Theoretical Methods.~}
We study the braiding of vortex core MZMs in a two-dimensional (2D) topological superconductor that arises from the interplay of an $s$-wave superconducting order parameter, a Rashba spin-orbit coupling, and the presence of ferromagnetism, and is described by the Hamiltonian
\begin{align}
\mathcal{H} =& \; -t_e \sum_{{\bf r, r'}, \beta} e^{{\text i}\theta_{\bf r r'}}c^\dagger_{{\bf r}, \beta} c_{{\bf r}^\prime, \beta} - \mu \sum_{{\bf r}, \beta} c^\dagger_{{\bf r}, \beta} c_{{\bf r}, \beta} \nonumber \\
     &+ \mathrm{i} \alpha \sum_{{\bf r}, \boldsymbol{\delta }, \beta, \gamma}  e^{{\text i}\theta_{\bf rr+\boldsymbol{\delta}}} c^\dagger_{{\bf r}, \beta} \left(\boldsymbol{\delta} \times \boldsymbol{\sigma} \right)^z_{\beta, \gamma} c_{{\bf r} + \boldsymbol{\delta}, \gamma} \nonumber \\
    & + \sum_{{\bf r}} \left( \Delta({\bf r}, t) c^\dagger_{{\bf r}, \uparrow} c^\dagger_{{\bf r}, \downarrow} + \Delta^*(
    {\bf r}, t) c_{{\bf r}, \downarrow} c_{{\bf r}, \uparrow} \right) \nonumber \\
    &+ JS \sum_{\bf r} \left( c^\dagger_{{\bf r}, \uparrow} c_{{\bf r}, \uparrow} - c^\dagger_{{\bf r}, \downarrow} c_{{\bf r}, \downarrow} \right) \; .
    \label{eq:H}
\end{align}
Here, the operator $c^\dagger_{{\bf r}, \beta}$ creates an electron with spin $\beta$ at site ${\bf r}$, $t_e$ is the nearest-neighbor hopping amplitude on a 2D square lattice, $\mu$ is the chemical potential, $\alpha$ is the Rashba spin-orbit (RSO) coupling between nearest-neighbor sites ${\bf r}$ and ${\bf r} +\boldsymbol{\delta}$, and $\Delta({\bf r},t)$ is the time and spatially dependent $s$-wave superconducting order parameter reflecting the motion of vortices. The hopping amplitude and the RSO coupling acquire a Peierls phase $\theta_{\bf rr'} = \frac{\pi}{\phi_0}\int_{\bf r}^{\bf r'} \bf A (\bf s)\cdot {\text d}{\bf s} $ arising from the vector potential $\bf A$ associated with the out-of-plane magnetic field necessary to induce vortices; each vortex corresponds to one flux quantum $\phi_0$ penetrating the system [for details, see Supplemental Material (SM) Sec.I].  The last term in Eq.~\ref{eq:H} describes the coupling between a local out-of-plane magnetic moment of magnitude $S$ at site ${\bf r}$ and the conduction electrons, with exchange coupling $J$. Such a Hamiltonian was proposed to describe the topological phase of FeSe$_{1-x}$Te$_x$ \cite{Mascot2022, Xu2023}, or those of magnet-superconductor hybrid systems, in which a layer of magnetic adatoms is placed on the surface of an $s$-wave superconductor \cite{Rachel2017,Palacio-Morales2019}. 
Given the ferromagnetic, out-of-plane alignment of the magnetic moments, we choose parameters such that the system is in a class D topological phase with Chern number $C=-1$ \cite{Rachel2017}. 

\begin{figure}
    \centering
    \includegraphics[width=0.9\columnwidth]{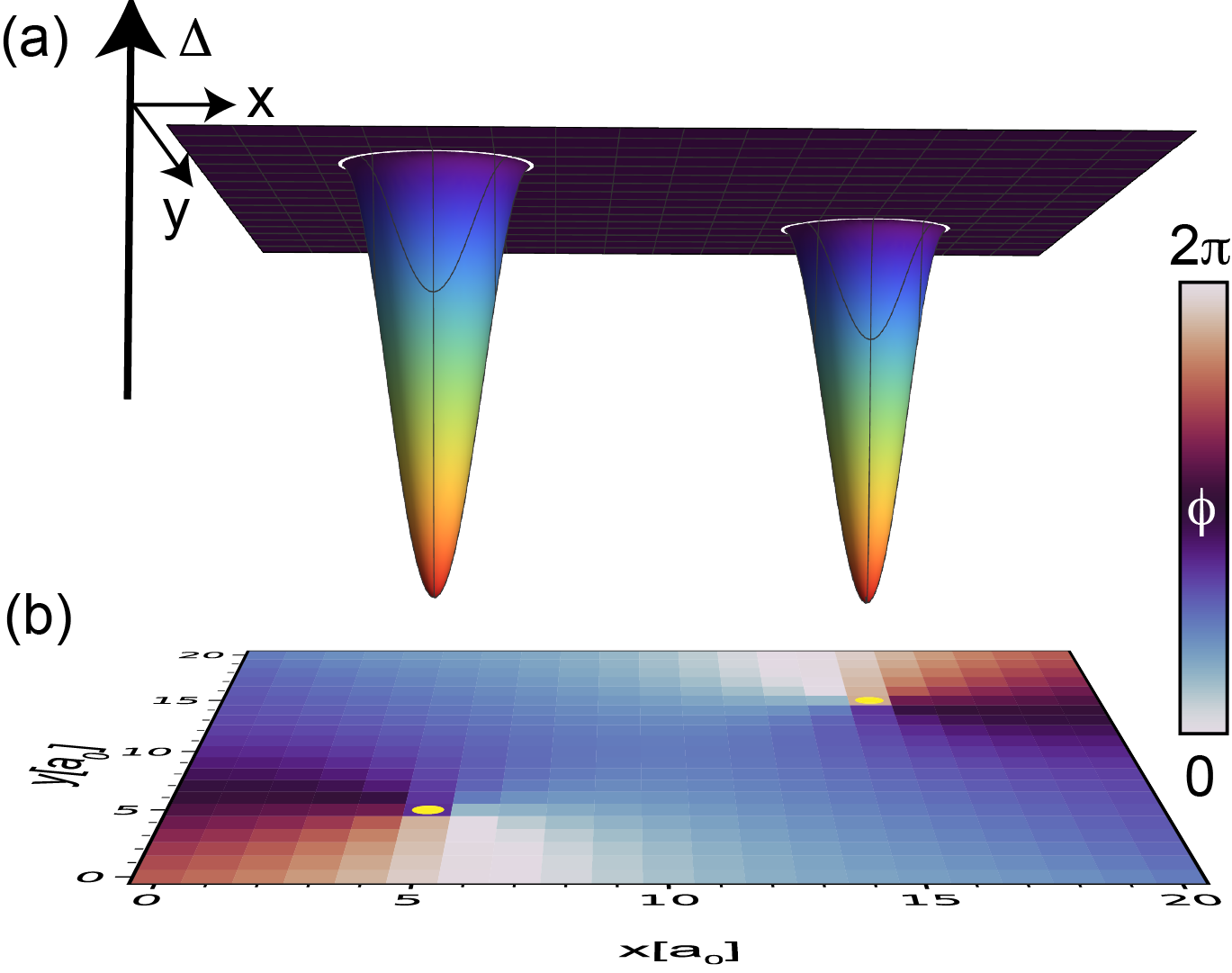}
    \caption{ Spatial form of the (a) magnitude and (b) phase of the superconducting order parameter $\Delta({\bf r}) = |\Delta({\bf r}) |e^{\text{i}\phi({\bf r})}$ for a system with two vortex cores.}
    \label{fig:Fig1}
\end{figure}
While the spatial structure of the superconducting order parameter is in general computed self-consistently using the Hamiltonian in Eq.(\ref{eq:H}) \cite{Vafek2001,Nagai2012,Smith2016}, this approach is computationally too demanding to perform for every time $t$ when the vortices are spatially moved. We therefore assume a time-independent local structure of the SCOP around a moving vortex core with time-dependent position ${\bf r}_\nu(t)=(x_\nu(t),y_\nu(t))$ which is given by 
$\Delta({\bf r}) = |\Delta({\bf r}) |e^{\text{i}\phi({\bf r})}$
where
\begin{equation}
    |\Delta({\bf r})| = \Delta_0 \sin^2\Big(\frac{\pi}{2R_V}|{\bf r}-{\bf r}_\nu|\Big) \ ,
\end{equation}
if ${\bf r}=(x,y)$ is within a distance of $R_V$ of ${\bf r}_\nu$, and $|\Delta({\bf r})|=\Delta_0$ otherwise, as shown in Fig.\ref{fig:Fig1}(a).
Moreover, the phase of the SCOP at site ${\bf r}$ is given by 
\begin{equation}
   \phi({\bf r}) = - \sum_{\nu=1}^{N_V} \sum_{i,j=-\infty}^\infty \arctan{\frac{y-y_\nu-jN_y}{x-x_\nu-iN_x}}
\end{equation}
where $N_V$ is the number of vortices in a supercell,  $N_x$ and $N_y$ is the supercell dimension in the $x$ and $y$ directions, and $i,j$ run over all copies of the supercell due to the periodic boundary conditions 
[see Fig.~\ref{fig:Fig1}(b)]. The time-dependence of the vortex cores' motion, and hence of the MZMs, is thus determined by ${\bf r}_\nu(t)$, with the time for the vortex cores to move between two lattice sites given by $t_V$ (for details, see the SM Sec.II).

\begin{figure}
    \centering
    \includegraphics[width=\columnwidth]{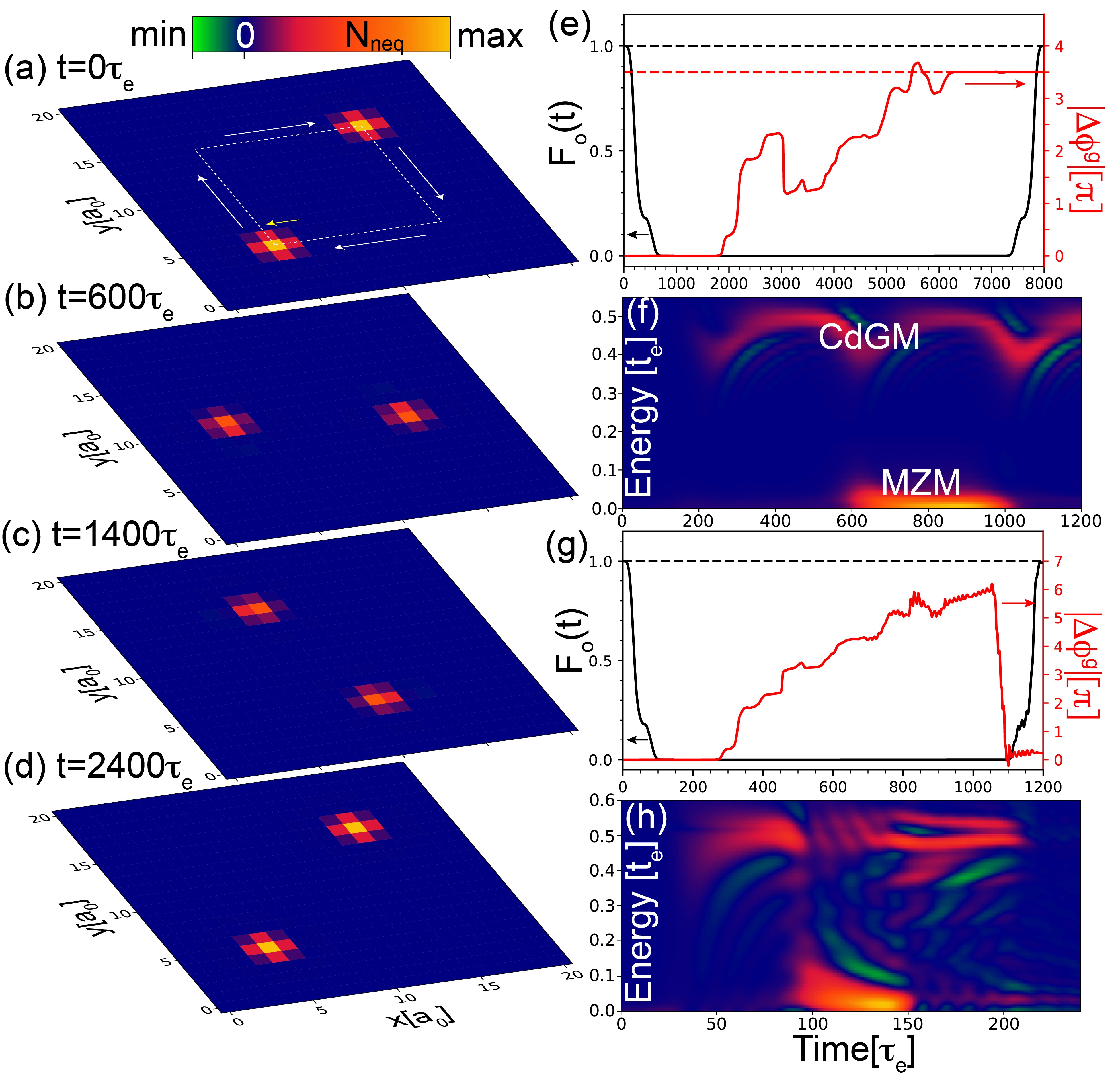}
    \caption{
    (a)-(d) Zero-energy $N_{\text{neq}}$ for different times during the $\sqrt{Z}$-gate process (the exchange path is shown by a dashed white line in (a)). Parameters are $(\mu, \alpha, \Delta_0, JS) = (-7, 0.9,2.4,4.4)t_e$. 
    (e) Fidelity and geometric phase difference for a $\sqrt{Z}$-gate with $t_{V}=400 \tau_e$. (f) $N_\mathrm{neq}$ as a function of time and energy at the site indicated by a yellow arrow in (a). (g),(h) same as (e) and (f) but with $t_{V}=60 \tau_e$. Parameters are $(\mu, \alpha, \Delta_0, JS, \Gamma) = (-4, 1.6,2.4,3.09,0.01)t_e$
    } 
    \label{fig:Fig2}
\end{figure}

The successful simulation of any quantum gate requires the process to be adiabatic in nature with a characteristic time scale for the vortex motion much larger than $\hbar/\Delta_t$, where $\Delta_t$ is the topological gap, thus avoiding any excitations between the MZMs and the trivial Caroli-de Gennes Matricon (CdGM) states inside the vortex core \cite{Caroli1964, Chen2018}. The adiabaticity is reflected in the time-dependent fidelity
\begin{equation}
    F_i(t) = |\langle\Psi_i(t)|\Psi_i(t=0)\rangle|
\end{equation}
where $\ket{\Psi_i(t)}$ ($i=e,o$) is the time-dependent many-body wave function of the system (for details of how $\ket{\Psi_i(t)}$ is constructed, see SM Sec.III) and $F_i(t)$ approaches unity at the end of an adiabatic process. Moreover, the anyon statistics of the vortex core MZMs is reflected in the geometric phase difference $\Delta \phi^g = \phi_e^g - \phi_o^g$ between the even- and odd-parity channels, with $\phi_{i}^g$ being obtained from the gauge- and parametrization-invariant functional \cite{Samuel1988,Mukunda1993}
\begin{equation}
\begin{aligned}
\phi_i^g(t) =& \arg \langle \Psi_i(t_0)| \Psi_i(t) \rangle - \text{Im}\int_{t_0}^t \left\langle 
 \Psi_i(t')|\partial_{t'}\Psi_i(t') \right\rangle \text{d}t'.
\end{aligned}
\end{equation}
The Ising anyon statistics of the MZMs require that $\Delta \phi^g$ be a half-odd-integer multiple of $\pi$ after a the execution of a $\sqrt{Z}$-gate, i.e., a single exchange of two MZMs. Finally, to visualize the gate operation in space and time, we employ the non-equilibrium local density of states $N_\text{neq}(t, \omega, {\bf r}) = -\frac{1}{\pi}\text{Im} \left[ G^\mathrm{r} (t,\omega,{\bf r}) \right]$ with the retarded Green's function matrix given by 
\begin{equation}
    \left[\text{i}\frac{\text{d}}{\text{d}t} + \omega + \text{i}\Gamma - \hat{H}(t) \right]\hat{G}^\mathrm{r}(t, \omega) = \hat{1} \; ,
\end{equation}
where $\hbar/\Gamma$ is the quasi-particle lifetime. It was previously shown \cite{Bedow2022} that $N_\text{neq}(t, \omega, {\bf r})$ is proportional to the time-dependent differential conductance measured in STM experiments. \\

{\it Results.~} 
{\it $\sqrt{Z}$-gate.~} To study the anyonic exchange statistics of the vortex core MZMs, we begin by investigating the form of the geometric phase under the exchange of two vortices, representing a $\sqrt{Z}$-gate. In Figs.~\ref{fig:Fig2}(a)-(d), we present the spatial form of the zero-energy $N_\text{neq}$ for several times during the gate process [the vortex path is indicated by a white dashed line in Fig.~\ref{fig:Fig2}(a), and the full time dependence of $N_\text{neq}$ is shown in Supplementary Movie 1]. In Fig.~\ref{fig:Fig2}(e), we plot the time-dependent fidelity and geometric phase difference for an adiabatic process, where the fidelity reaches unity and $\Delta \phi^g = 7\pi/2$ after the conclusion of the exchange process, thus confirming the anyonic nature of the vortex core MZMs. In Fig.~\ref{fig:Fig2}(f), we present the non-equilibrium LDOS $N_\text{neq}$ as a function of time and energy at a site in the exchange path marked by the yellow arrow in Fig.~\ref{fig:Fig2}(a).
We find that as the vortex core MZM moves through this site, the MZM remains well separated from the trivial CdGM states inside the vortex core, thus reflecting the adiabaticity of the process. However, as we decrease the characteristic time for the vortex motion, $t_V$, and though the fidelity is only slightly reduced to 0.99 [see Fig.~\ref{fig:Fig2}(g)], the geometric phase difference $\Delta \phi^g$ already deviates significantly from the expected value of an odd multiple of $\pi/2$, indicating the loss of adiabaticity. This is also reflected in the time and energy-dependent $N_\text{neq}$, shown in Fig.~\ref{fig:Fig2}(h), which in contrast to the results shown in Fig.~\ref{fig:Fig2}(f), now reveals a strong hybridization between the MZM and the trivial CdGM states, leading to the loss of  adiabaticity.\\

\begin{figure}
    \centering
    \includegraphics[width=\columnwidth]{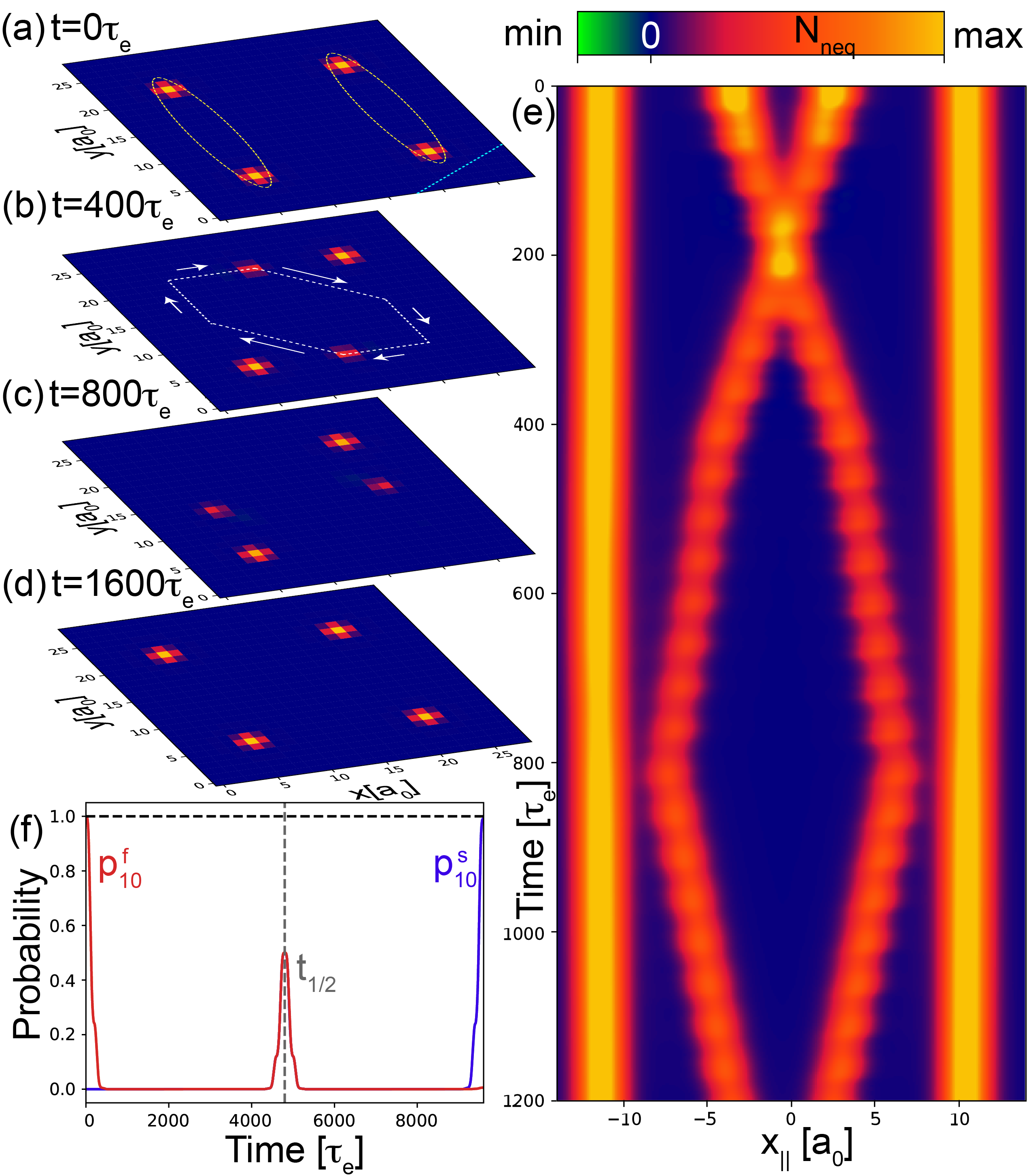}
    \caption{(a)-(d) Zero-energy $N_\mathrm{neq}$ for various times during the execution of a $\sqrt{X}$-gate with $t_V=50\tau_e$. Dashed white line and arrows indicate the path and direction in which the vortices are moved. (e) Majorana world lines representing a $\sqrt{X}$-gate obtained by projecting the zero-energy $N_\mathrm{neq}$ onto an axis shown as a yellow dashed line in (a). (f) Time-dependent success and fail probabilities during the execution of the $X$-gate with $t_V=200\tau_e$. $t_{1/2}$ indicates the time when the MZMs have been exchanged once, realizing a $\sqrt{X}$-gate.  Parameters are $(\mu, \alpha, \Delta_0, JS, \Gamma) = (-7, 0.9, 2.4, 4.4, 0.01)t_e$.} 
    \label{fig:Fig3}
\end{figure}

{\it $\sqrt{X}$- and $X$-gates.~}
To implement a one-qubit $X$-gate, we consider an MSH system with 2 pairs of vortices, as shown in Fig.~\ref{fig:Fig3}(a). The two logical states of the one-qubit system can be formed by either the even-parity states, denoted by $\ket{00}$ and $\ket{11}$ in Fock notation, respectively, or by the odd-parity states $\ket{01}$ and $\ket{10}$ \cite{Bedow2024}. Here, the first (second) entry in $\ket{ij}$ ($i,j=0,1$) describes the occupation of the first (second) pair of vortices [the pairs are indicated by yellow dashed lines in Fig.~\ref{fig:Fig3}(a)], resulting in either the vacuum $(0)$ or an electron $(1)$ after fusion. Below, we consider the $X$-gate in the odd-parity sector, as the energy difference between the $\ket{01}$ and $\ket{10}$ states, arising from the finite size of the system and the resulting hybridization between the MZMs, can be nearly reduced to zero in this sector, thus avoiding spurious dynamic phase effects on the $X$-gate process \cite{Bedow2024}.

The $X$-gate is executed by exchanging two vortices, each belonging to one of the two pairs, twice. The zero-energy $N_\mathrm{neq}$ for various times during the gate process are shown in Figs.~\ref{fig:Fig3}(a)-(d) (the full time dependence of $N_\mathrm{neq}$ is shown in Supplementary Movie 2). The exchange path, represented by a white dashed line in Fig.~\ref{fig:Fig3}(b), was chosen such that, given the computational constraints on the size of the system, the hybridization between the MZMs is minimized at any point during the gate process. 
To ascertain the success of the $X$-gate execution, we consider the time-dependent many-body wave function of the system $\ket{10(t)}$, with initial state $\ket{10(t=0)} = \ket{10}$. Since $X\ket{10}=\ket{01}$, we can define the time-dependent success probability for the execution of the $X$-gate via $p^s_{10}(t)=|\braket{01|10(t)}|^2$, and a fail probability via $p^f_{10}(t)=|\braket{10|10(t)}|^2$; a perfect execution of the $X$-gate completed at time $t=t_f$ then yields $p^s_{10}(t_f)=|\braket{01|10(t_f)}|^2=1$ and $p^f_{01}(t_f)=0$.
In Fig.~\ref{fig:Fig3}(f), we present $p^s_{10}(t)$ and $p^f_{10}(t)$ during the execution of the $X$-gate. After the first exchange of two vortices at $t=t_{1/2}$, the half way point of the $X$-gate, we obtain $p^s_{10}(t_{1/2})=0.498$ and $p^f_{10}(t_{1/2})=0.5$, representing a successful implementation of a $\sqrt{X}$-gate, since $\sqrt{X}\ket{10}=\left(\ket{10}+\ket{01}\right)/\sqrt{2}$. Similarly, after the conclusion of the entire $X$-gate process, we have $p^s_{10}(t_f)=0.991$ and $p^f_{10}(t_{f})=0.006$, demonstrating the successful realization of the $X$-gate, as the errors are within the error correction threshold \cite{Raussendorf2007,Stace2009}. To visualize the gate process in time and space, we present in Fig.~\ref{fig:Fig3}(f) the Majorana world lines for the $\sqrt{X}$-gate -- reflecting a single exchange of 2 MZMs -- obtained by projecting the zero-energy $N_\mathrm{neq}$ onto the dashed blue line shown in Fig.~\ref{fig:Fig3}(a). Finally, we note that the successful simulation of the $\sqrt{Z}$-, $\sqrt{X}$-, and $X$-gates demonstrated above implies that all Clifford gates can be simulated using vortex core MZMs. 

\begin{figure}
    \centering
    \includegraphics[width=1\linewidth]{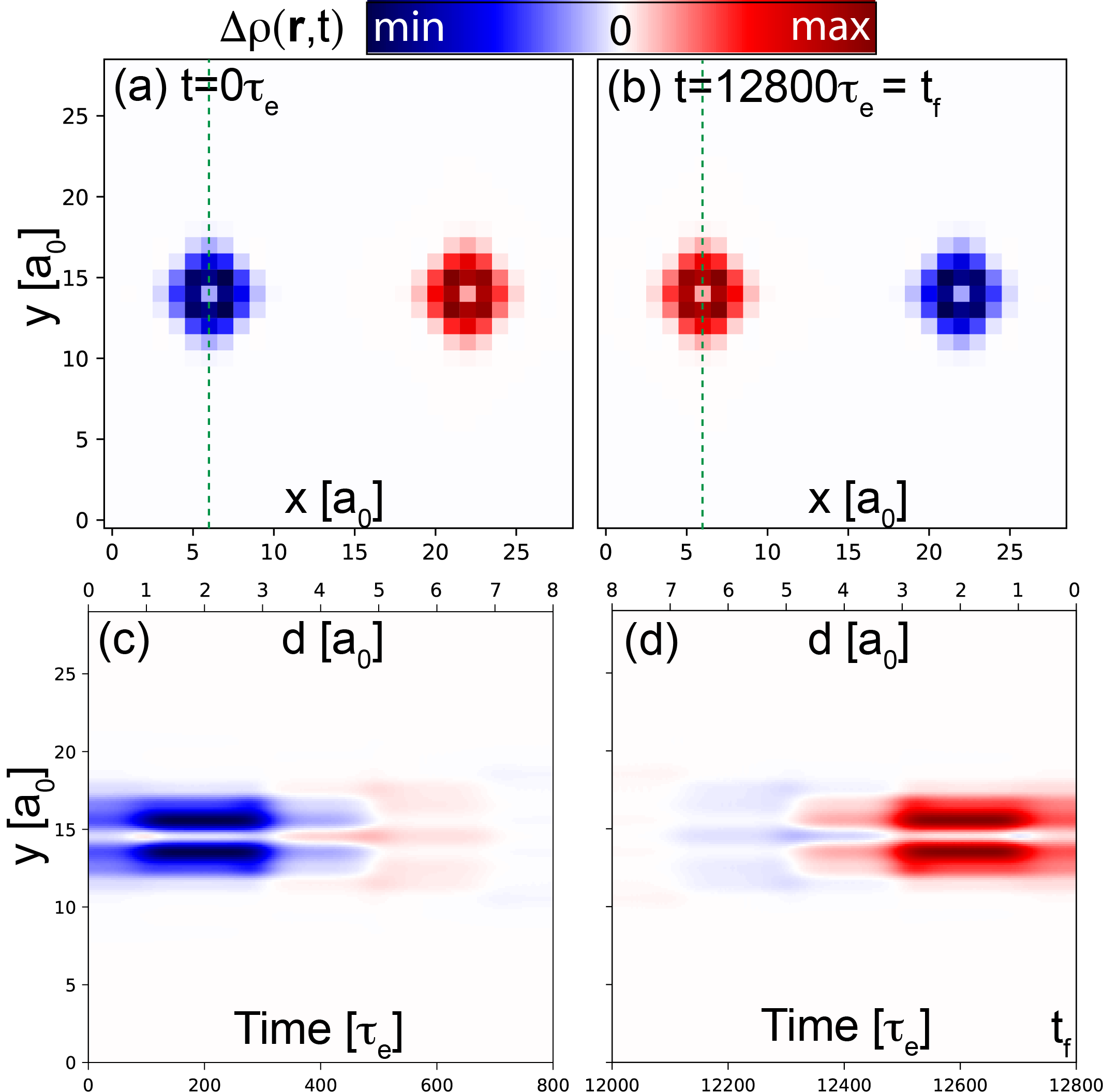}
    \caption{
   Charge density difference between the odd-parity $|10\rangle$ and $|01\rangle$ states when the vortices are fused at (a) $t=0$ and (b) $t=t_f$. Time-dependent charge density difference [along the green dotted line in (a)] (c) when the vortices are initially moved apart after $t=0$, and (d)  when the vortices are fused again at $t_f$. Parameters are $(\mu, \alpha, \Delta_0, JS) = (-7, 0.9, 2.4, 4.4)t_e$, with $t_V = 200 \tau_e$.}
    \label{fig:Fig4}
\end{figure}

{\it Readout of the qubit state.~}
It was recently shown that MZM qubit states can be experimentally read out based on the charge density resulting from the fusion of an MZM pair \cite{Hodge2025_2,Bedow2025,Pandey2025}. 
This is achieved by moving the vortices belonging to the same MZM pair towards each other (it was suggested that the vortex-vortex repulsion can be overcome using AFM techniques \cite{November2019}).  The fusion rule $\gamma \times \gamma = 1 + \psi$, where $\gamma$ represents a MZM, $1$ the vacuum, and $\psi$ an electron, implies that fusion of a single MZM pair leads to a charge difference between the $\ket{0}$ and $\ket{1}$ states, which is quantized to $-e$ in the limit of small SCOP \cite{Hodge2025_2}. Thus, to read out the quantum state of the many-body system, $|ij(t)\rangle$ with $i,j=0,1$, we determine the occupation of the MZM pairs by computing the charge density using
\begin{equation}
    \rho_{ij}({\bf r}, t) =  -e \sum_{\sigma=\uparrow, \downarrow} \braket{ij(t)|c^\dagger_{{\bf r}, \sigma} c_{{\bf r}, \sigma}|ij(t)} \; .
\end{equation}
To emphasize the difference in the fusion outcome between two qubit states, we consider below the charge density difference 
\begin{equation}
    \Delta \rho ({\bf r}, t) = \rho_{10}({\bf r}, t) -\rho_{01}({\bf r}, t) \ .
\end{equation}
To best demonstrate how the action of the $X$-gate on the qubit states affects the charge density, we consider symmetric start and end points of the gate process, when each pair of vortices is fused. Thus, we start with fused vortices at $t=0$, then move them apart to bring them into the spatial locations shown in Fig.~\ref{fig:Fig3}(a), subsequently perform the $X$-gate, and finally fuse them again at $t=t_f$. In Figs.~\ref{fig:Fig4}(a) and (b), we present $\Delta \rho ({\bf r}, t)$ at $t=0$ and $t=t_f$, respectively. Since $X\ket{10}=\ket{01}$ and $X\ket{01}=\ket{10}$, we expect $\Delta \rho ({\bf r}, t=t_f)=-\Delta \rho ({\bf r}, t=0)$, which is confirmed by the numerically computed charge density difference shown in Figs.~\ref{fig:Fig4}(a) and (b). We note that the charge density remains stable after the end of the fusion process, as the vortices represent topological defects that trap the charge; this will facilitate the detection of the charge using capacitive experimental measurements. Moreover, it is interesting to note that a non-zero charge density only emerges when the MZMs are sufficiently close such that they hybridize. This is reflected in the time-evolution of the charge density near $t=0$ shown in Fig.~\ref{fig:Fig4}(c) along the line-cut indicated in Fig.~\ref{fig:Fig4}(a); the charge density difference vanishes as the vortices, and hence the MZMs, are moved further apart (the distance between vortices is shown on the upper x-axis).  
Conversely, we find that as $t$ approaches $t_f$ [see Fig.~\ref{fig:Fig4}(d)], a non-zero charge density appears only when the vortices are brought sufficiently close together (the spatially and energy resolved $N_\mathrm{neq}$ during the fusion process is shown in Supplemental Movie 3). Our results thus demonstrate that the qubit's many-body state can be read out by fusing the vortex pairs, and measuring the resulting charge distribution. 

{\it Discussion} We have demonstrated that topologically protected $\sqrt{Z}$-, $\sqrt{X}$- and $X$-gates can be successfully simulated in topological superconductors of class D using MZMs localized in the magnetic vortex cores.  To this end, we computed the geometric phase difference and transition probabilities between qubit states which directly reflected the anyonic nature of the MZMs, as well as their non-Abelian braiding statistics. In addition to the MZMs located in 1D magnet-superconductor hybrid networks \cite{Bedow2024,Bedow2025}, the vortex core MZMs thus provide an suitable platform for the realization of fault-tolerant quantum computing. Proposals for the braiding of such vortices using atomic force microscopy (AFM) techniques \cite{November2019} represent an exciting new approach to the implementation of the above discussed protocols.   \\   

{\it Acknowledgments}
This work was supported by the U.\ S.\ Department of Energy, Office of Science, Basic Energy Sciences, under Award No.\ DE-FG02-05ER46225. We wish to thank J. Hoffman, E. Mascot, and C. Xu for stimulating discussions.

\end{document}